\pdfoutput=1

\documentclass[11pt]{article}

\usepackage[final]{acl}

\usepackage{times}
\usepackage{latexsym}
\usepackage{amsmath}
\usepackage{amssymb}
\usepackage{multirow}
\usepackage[normalem]{ulem}
\useunder{\uline}{\ul}{}
\usepackage{titlesec}

\titleformat{\subsection}{\normalsize\bfseries}{\thesubsection}{1em}{}

\titleformat{\subsubsection}{\fontsize{10.7pt}{11pt}\bfseries\selectfont}{\thesubsubsection}{1em}{}

\usepackage[T1]{fontenc}

\usepackage[utf8]{inputenc}

\usepackage{microtype}

\usepackage{inconsolata}

\usepackage{graphicx}

\title{DeepRTL2: A Versatile Model for RTL-Related Tasks}

\author{
Yi Liu$^{1,2}$\thanks{These authors contributed equally.}, 
Hongji Zhang$^{1,2}$\footnotemark[1], 
Yunhao Zhou$^{1,2}$, \\
\textbf{Zhengyuan Shi$^{1,2}$,} 
\textbf{Changran Xu$^{1,2}$,} 
\textbf{Qiang Xu$^{1,2}$} \\
$^1$The Chinese University of Hong Kong \\
$^2$National Technology Innovation Center for EDA \\
\texttt{\{yliu22,zyshi21,qxu\}@cse.cuhk.edu.hk} \\
\texttt{\{hongjizhang183,yunhaoz.cs,xxuchangran\}@gmail.com}
}

\begin{document}
\maketitle
\begin{abstract}
The integration of large language models (LLMs) into electronic design automation (EDA) has significantly advanced the field, offering transformative benefits, particularly in register transfer level (RTL) code generation and understanding. While previous studies have demonstrated the efficacy of fine-tuning LLMs for these generation-based tasks, embedding-based tasks, which are equally critical to EDA workflows, have been largely overlooked. These tasks, including natural language code search, RTL code functionality equivalence checking, and performance prediction, are essential for accelerating and optimizing the hardware design process. To address this gap, we present \textbf{DeepRTL2}, a family of versatile LLMs that unifies both generation- and embedding-based tasks related to RTL. By simultaneously tackling a broad range of tasks, DeepRTL2 represents the first model to provide a comprehensive solution to the diverse challenges in EDA. Through extensive experiments, we show that DeepRTL2 achieves state-of-the-art performance across all evaluated tasks.


\end{abstract}
\section{Introduction}
\label{sec_introduction}
The rapid advancement of large language models (LLMs) has had a profound impact on various domains~\citep{singhal2023large,bran2024transformers}, including electronic design automation (EDA). 
Recently, LLMs have shown remarkable potential in automating and enhancing tasks related to the generation and understanding of register transfer level (RTL) code~\citep{liu2024rtlcoder,zehuabetterv,zhao2024codev,liu2025deeprtl}.
These models are capable of generating RTL code from high-level natural language instructions or summarizing the functionality of existing RTL code, thereby substantially improving the efficiency of hardware design workflows.
While the application of LLMs to these generation-based tasks has yielded impressive results, their full potential at the RTL stage remains underexplored, particularly in embedding-based tasks that are equally crucial to the design process.

Embedding-based tasks like natural language code search, RTL code functionality equivalence checking, and performance prediction are vital for accelerating and optimizing the hardware design process. 
Natural language code search allows designers to quickly query large RTL codebases with simple natural language descriptions, enabling efficient identification and reuse of relevant modules, thus reducing search time.
Moreover, verification and optimization are two key time-consuming bottlenecks in hardware design. RTL code functionality equivalence checking can significantly reduce the time spent on verification by quickly assessing whether two designs are functionally equivalent.
Performance prediction tasks, such as power, performance, and area (PPA) estimation, enable early evaluation of RTL design efficiency. Accurate performance predictions can guide RTL code optimization, minimizing the need for time-intensive trial-and-error.
Together, these tasks enhance code reuse, verify functionality, and provide early performance feedback, resulting in a more streamlined and efficient design workflow.
Previous methods have attempted to apply machine learning solutions for hardware design verification~\citep{vasudevan2021learning} and performance prediction~\citep{fang2023masterrtl}, but they are typically design-specific, lack generalizable representations for RTL designs, or do not operate directly at the RTL stage.

\begin{figure*}[h]
    \centering
    \includegraphics[width=0.9\linewidth]{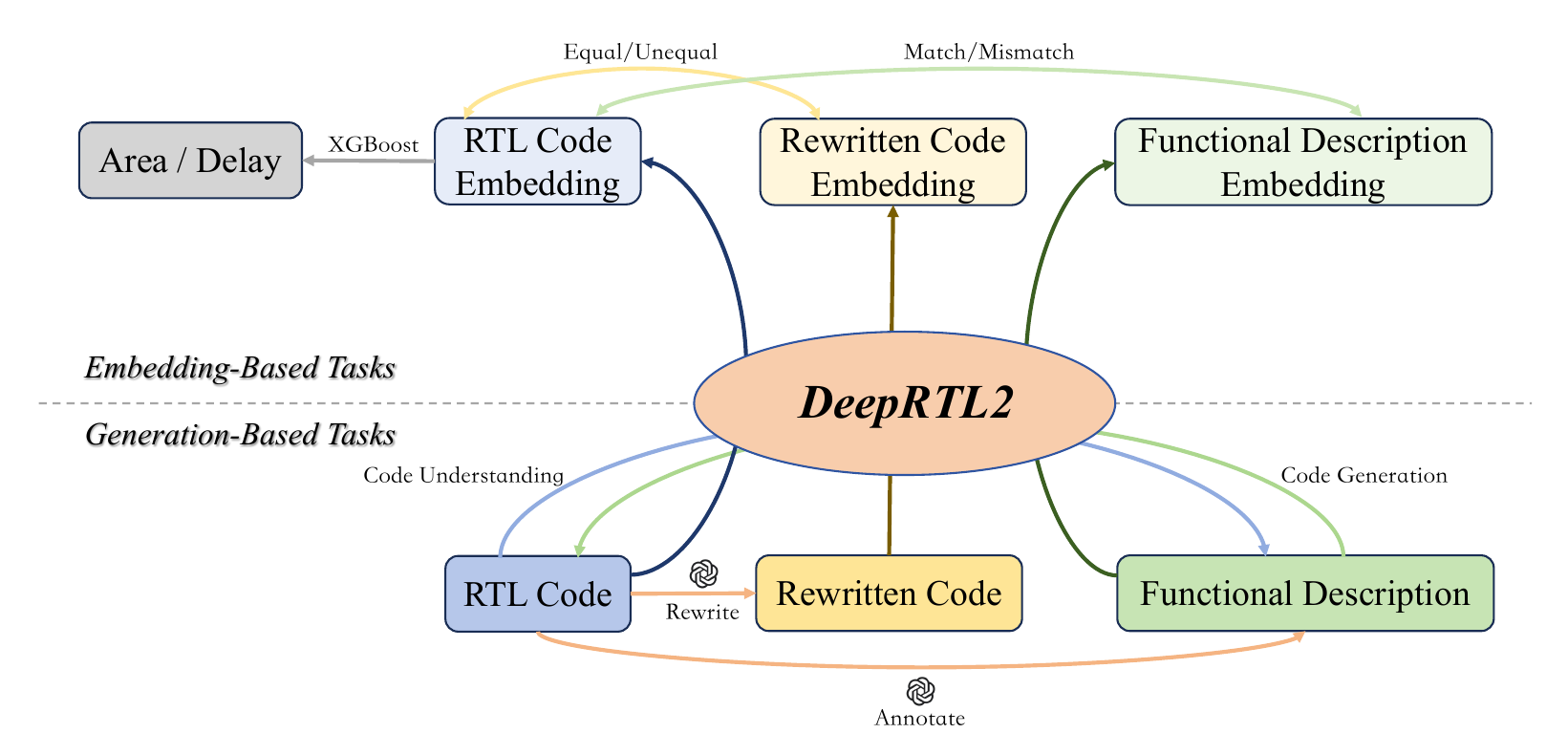}
    \vspace{-12pt}
    \caption{The overview of DeepRTL2. It can handle both generation- and embedding-based tasks at the RTL stage. For generation-based tasks, it performs RTL code generation and understanding. For embedding-based tasks, it uses cosine similarity scores between the embeddings of RTL code and functional descriptions to assess their match, enabling natural language code search. Additionally, cosine similarity between RTL code embeddings and rewritten code embeddings is used for functionality equivalence checking. Furthermore, a prediction model, such as XGBoost, can be applied to predict area and delay metrics based on code embeddings.}
    \vspace{-12pt}
    \label{fig:overview}
\end{figure*}

In this paper, we introduce \textbf{DeepRTL2}, a family of versatile LLMs designed to address both generation- and embedding-based tasks related to RTL. By unifying these tasks in a single model, DeepRTL2 offers a comprehensive solution to the multifaceted challenges inherent in EDA. Unlike previous work, which has primarily focused on generation, DeepRTL2 is the first model to provide a unified framework for handling a broad range of critical EDA tasks, including code generation, understanding, natural language code search, functionality equivalence checking, and performance prediction.
Figure~\ref{fig:overview} provides an overview of our model.
To achieve this, we have carefully curated a comprehensive dataset and developed new benchmarks for each task, with a particular focus on the embedding-based tasks, for which no existing datasets or benchmarks are available.
We have adopted state-of-the-art decoder-only models, such as Llama-3.1~\citep{dubey2024llama} and DeepSeek-Coder~\citep{guo2024deepseek}, as our base models for fine-tuning, given their superior performance over other architectures in the open-source LLM space.
To enable these models to handle both generation- and embedding-based tasks, we adapt the generative representational instruction tuning (GRIT) approach~\citep{muennighoff2025generative} for fine-tuning, ensuring that DeepRTL2 can effectively manage the diverse tasks at the RTL stage.
Through extensive experimentation, we demonstrate that the DeepRTL2 series achieves state-of-the-art performance across all evaluated tasks.
\section{Related Works}
\label{sec_related_work}
\subsection{Register Transfer Level in EDA}
Register transfer level (RTL) is a key abstraction in EDA that describes the flow of data between registers and the operations performed on this data. It is typically expressed using hardware description languages (HDLs), with Verilog being the most widely used HDL in the industry. Thus, throughout this paper, we use the terms RTL code and Verilog code interchangeably. 
In modern hardware design, engineers usually begin with specifications in natural language, which are then manually translated into HDLs before synthesizing the circuit elements~\citep{blocklove2023chip}. 
RTL serves as an intermediary between high-level design specifications and low-level implementation details, enabling designers to describe intricate digital systems while retaining flexibility for synthesis into gate-level representations.
Within EDA workflows, RTL plays a crucial role in various phases, including functional verification, performance estimation, synthesis, and optimization. Efficient handling of RTL code is essential for minimizing design time, improving performance, and ensuring correctness.

\subsection{LLMs for RTL}

With the rapid development of artificial intelligence (AI), there has been increasing interest in leveraging these technologies to automate and enhance RTL-based design workflows~\citep{chen2024large}.
A key area of focus has been the use of LLMs for RTL code generation and understanding, which has shown great promise in improving hardware design efficiency~\citep{thakur2023benchmarking,liu2023verilogeval,lu2024rtllm}. 
Recent works have fine-tuned open-source LLMs to generate high-quality RTL code from natural language descriptions~\citep{chang2024data,liu2024rtlcoder,thakur2024verigen,zehuabetterv,zhang2024mg,cui2024origen,zhao2024codev}, achieving significant improvements in the automation of hardware design process. Additionally, models like DeepRTL~\citep{liu2025deeprtl} have extended these capabilities by introducing RTL code understanding tasks, \emph{i.e.}, summarizing the functionality of existing code, which facilitates collaboration and comprehension among hardware designers.
Despite the great success achieved in these generation-based tasks, prior research has largely overlooked embedding-based tasks, which are equally critical for addressing challenges in EDA.
Embedding-based tasks, such as natural language code search, RTL code functionality equivalence checking, and performance prediction, are essential for improving the efficiency of code reuse, verification, and optimization within hardware design workflows.
Unlike generation-based tasks, which focus on producing new RTL code, embedding-based tasks involve understanding and analyzing existing designs, providing valuable insights into design reusability, correctness, and performance.
Meanwhile, even if some studies have applied machine learning techniques for hardware design verification~\citep{vasudevan2021learning} and performance prediction~\citep{fang2023masterrtl}, these efforts are either design-specific, lack generalizable representations for RTL designs, or do not operate directly at the RTL stage.
In contrast, this work introduces DeepRTL2, a versatile model capable of handling both generation- and embedding-based tasks, achieving superior performance across all evaluated tasks despite its versatility.

\subsection{Embedding Capabilities of Decoder-Only LLMs}
Compared to bidirectional encoders like BERT~\citep{devlin2018bert} and encoder-decoder architectures like T5~\citep{raffel2020exploring}, decoder-only LLMs have demonstrated superior performance across a range of language tasks~\citep{brown2020language}. 
However, their potential for text embedding tasks was largely overlooked until recently. 
In recent years, several studies have focused on adapting decoder-only LLMs for language embedding tasks~\citep{jiang2023scaling,wang2023improving,behnamghader2024llm2vec,springer2024repetition,lei2024meta,lee2024nv}.
Notably, \citet{muennighoff2025generative} introduce the GRIT training strategy, which employs a multi-task training objective function to enable a single decoder-only LLM to both generate content and encode text into fixed-length vectors. 
Despite their success on various language embedding benchmarks, these models primarily focus on general embedding tasks, which limits their effectiveness on specialized tasks like RTL embedding-based tasks.
To the best of our knowledge, there is no model that has been specifically trained for RTL embedding, despite its critical role in optimizing hardware design workflows. DeepRTL2 is the first model explicitly designed for RTL embedding-based tasks, outperforming general-purpose text embedding models on our benchmarks.
\section{Dataset}
\label{sec_dataset}

Previous research has predominantly focused on generation-based tasks, resulting in a notable gap in available datasets for the embedding-based tasks considered in this paper. 
Moreover, the availability of RTL code is limited even for generation-based tasks, due to the proprietary nature of hardware designs. 
To fill this gap, we have curated a comprehensive dataset tailored to support both generation- and embedding-based tasks at the RTL stage. 
Furthermore, we have established new benchmarks specifically for the embedding-based tasks, which have been largely neglected in previous work.

\subsection{Generation-Based Tasks}

\subsubsection{RTL Code Generation}
\label{subsubsection:rtl_code_generation}

RTL code generation involves automatically synthesizing RTL code from user-defined natural language descriptions, streamlining hardware design and enabling a more accessible development process.
To construct a high-quality dataset for this task, we follow the data construction pipeline proposed in DeepRTL~\citep{liu2025deeprtl}, given its demonstrated effectiveness in generation-based tasks. 
The process begins by collecting \texttt{.v} files from GitHub\footnote{\url{https://github.com/}} using the keyword \texttt{Verilog}. 
\begin{figure}[h]
    \centering
    \includegraphics[width=\linewidth]{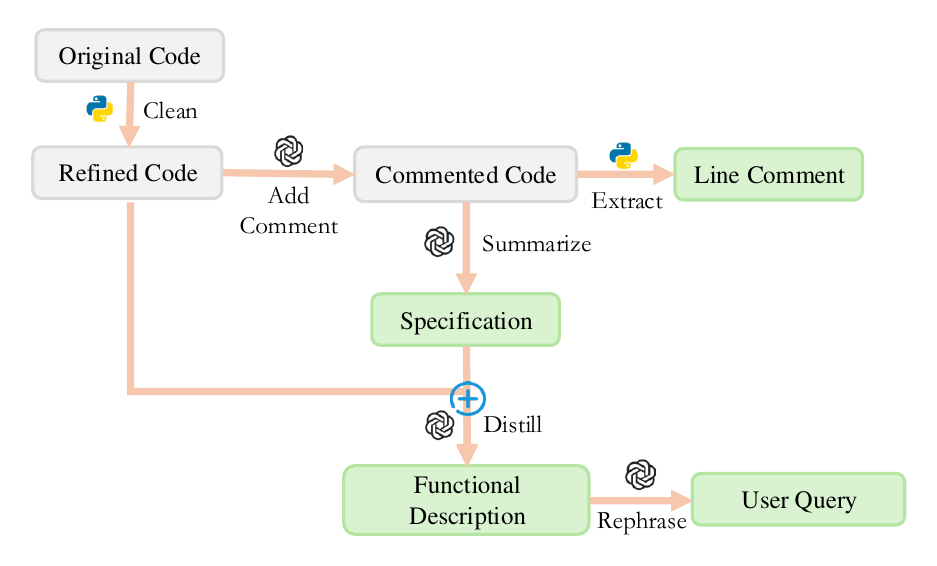}
    \vspace{-20pt}
    \caption{The annotation process for the RTL code generation/understanding dataset. After obtaining the high-level functional description, we prompt GPT-4o to rephrase it into a user query format, which is then used to construct the natural language code search dataset.}
    \vspace{-12pt}
    \label{fig:annotation_generation}
\end{figure}
Each file is then segmented into individual Verilog modules, with each module representing a distinct functional unit. 
To ensure dataset quality and reduce redundancy, we remove modules that are predominantly composed of comments or lack structurally complete \texttt{module} and \texttt{endmodule} declarations. 
Additionally, we apply MinHash and Jaccard similarity metrics~\citep{yan2017privmin} to eliminate duplicates. 
To further refine the dataset, we employ the Stagira Verilog parser~\citep{chen2023incremental} to filter out modules containing syntax errors, ensuring that only syntactically valid Verilog code is retained.

For annotation, we adopt the chain-of-thought (CoT) prompting strategy used in DeepRTL, leveraging GPT-4o~\citep{hurst2024gpt}, a state-of-the-art LLM, to generate structured and informative annotations. 
Specifically, we first query GPT-4o to insert line-level comments into the Verilog modules, then extract line-level descriptions, pairing individual lines of RTL code with corresponding natural language explanations.
Next, we prompt GPT-4o to generate a detailed specification for each module, comprising a summary of the module's functionality and a comprehensive explanation of its implementation process. 
By integrating these specifications with the module code, we construct high-level functional descriptions—succinct one-sentence summaries that capture the core functionality of each Verilog module.
The resulting dataset consists of Verilog modules enriched with line-level comments, detailed specifications, and succinct high-level functional descriptions, facilitating both generation and understanding tasks in RTL design.
Figure~\ref{fig:annotation_generation} provides an overview of the annotation process. For specifics on the prompts used, please refer to DeepRTL.
To ensure the quality of the generated annotations, we have conducted human evaluations, as detailed in Appendix~\ref{appendix:human_evaluation_for_generated_annotations}.

To further expand the training dataset and improve model performance, we augment our dataset with open-source Verilog datasets from RTLCoder~\citep{liu2024rtlcoder}, MG-Verilog~\citep{zhang2024mg}, and DeepCircuitX~\citep{li2025deepcircuitx}. These datasets provide additional RTL designs with diverse structures and functionalities, while also incorporating different annotation strategies. The diversity in annotations improves the model’s adaptability to varying description styles, enhancing its robustness across various RTL-related tasks.

\subsubsection{RTL Code Understanding}
\label{subsubsection:rtl_code_understanding}
RTL code understanding focuses on summarizing the functionality of existing Verilog code, enhancing collaboration and comprehension among hardware designers. 
The dataset for this task is derived from the RTL code generation dataset, with Verilog code as input and corresponding natural language description as output.
In the absence of a standardized benchmark for this task, we build upon the benchmark introduced in DeepRTL, which originally comprises 100 Verilog designs. To improve evaluation reliability and ensure broader coverage, we extend this benchmark to include 500 high-quality Verilog modules with diverse functionalities.
Each module is annotated by professional hardware designers with a concise summary of its functionality along with a detailed description of the specific operations involved in its execution.
This extended benchmark establishes a more robust and comprehensive foundation for evaluating RTL code understanding capabilities.

\subsection{Embedding-Based Tasks}

\subsubsection{Natural Language Code Search}\label{data:nlcs}

Natural language code search refers to the process of querying a large codebase using natural language to find relevant code snippets. 
It involves embedding both the user query and each code snippet into vectors, then calculating their similarity. The snippet with the highest similarity score is considered the best match for the user's requirements.
This task is particularly crucial for hardware design, as it enables code reuse, improves efficiency, and accelerates the transition from user specifications to RTL code.
For this task, we reuse the dataset and benchmark from the RTL code understanding task.
However, since the functional descriptions in the understanding dataset often contain specific identifiers, introducing the risk of data leakage, and are too complex for direct use in practical code search, we employ GPT-4o to rephrase the descriptions into a user query format, as shown in Figure~\ref{fig:annotation_generation}.
The rephrasing ensures that the new descriptions meet the following conditions: (1) no references to specific identifiers, (2) retention of the core functionality and high-level logic, and (3) clarity and simplicity, resembling how a user would query for relevant code based on its functionality.
\begin{figure}[ht]
    \centering
    \includegraphics[width=\linewidth]{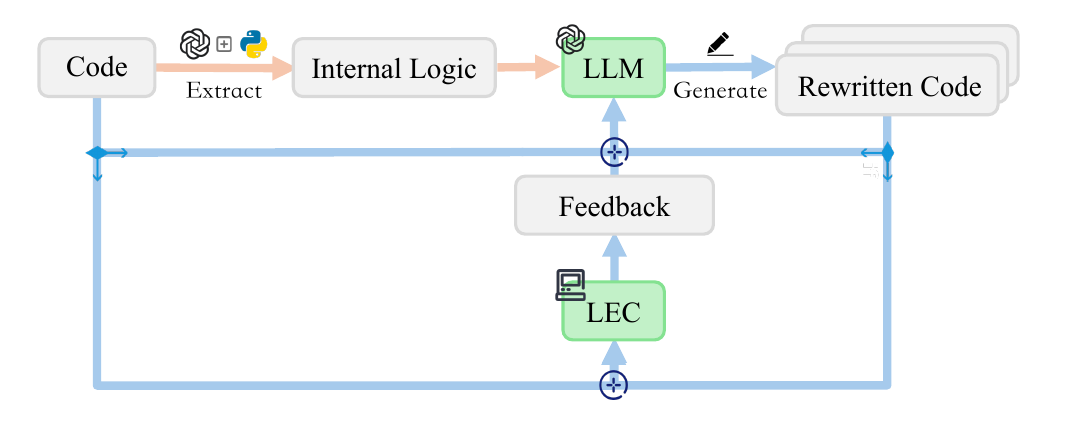}
    \vspace{-12pt}
    \caption{The feedback-driven code rewrite process.}
    \vspace{-12pt}
    \label{fig:code_rewrite_process}
\end{figure}
After this rephrasing process, we obtain the natural language code search dataset and benchmark in the format $\{(\text{user\_query}_i, \text{RTL\_code}_i)\}_{i=1}^n$. For details on the prompt used to rephrase the functional descriptions, please refer to the Appendix~\ref{appendix:prompt_for_rephrasing_descriptions}.

\subsubsection{Functionality Equivalence Checking}\label{data:fec}
\vspace{-6pt}

Functionality equivalence checking is a critical verification step in hardware design, ensuring that different RTL implementations exhibit identical behavior despite structural differences. 
To construct a dataset for this task, we develop a feedback-driven CoT prompting strategy using GPT-4o, as shown in Figure~\ref{fig:code_rewrite_process}.
Given a Verilog module, we first prompt GPT-4o to introduce significant modifications to its internal logic while preserving its intended functionality. 
We then use Yosys~\citep{yosys} to perform logic equivalence checking (LEC), which verifies whether the original and modified designs are functionally equivalent. 
Based on Yosys feedback—classified as equivalent, inequivalent, or syntax error—we iteratively refine the modifications. Specifically, we incorporate the original design, rewritten design, and verification results into the prompt to guide GPT-4o in generating alternative implementations.
This process is repeated for two to three rounds per design, ensuring a diverse set of functionally equivalent and inequivalent pairs.
The resulting dataset consists of paired RTL designs, where some maintain functional equivalence while others introduce subtle variations. 
Since only implementation details differ, distinguishing equivalent from inequivalent designs presents a significant challenge for models. 
Additionally, we adapt RTLLM v2.0~\cite{lu2024rtllm}, a Verilog generation benchmark, to construct a new benchmark for functionality equivalence checking. Applying the same feedback-driven CoT strategy to its 50 verified Verilog designs, we generate multiple alternative implementations, expanding our benchmark to 400 code pairs. This benchmark provides a diverse and well-validated resource for evaluating functionality equivalence checking. For further details on this process, please refer to the Appendix~\ref{appendix:code_rewrite_instructions}.

\subsubsection{Performance Prediction}\label{data:pp}

Performance prediction plays a crucial role in the early stages of hardware design, enabling designers to estimate key circuit characteristics before physical implementation. 
Accurate predictions allow for informed architectural decisions, reducing design iterations and improving overall efficiency.
Among the commonly used PPA metrics, delay and area are the primary focus in early-stage evaluations, as accurate power estimation requires detailed workload to specify the circuit's dynamic behavior, which is unavailable at the RTL stage.
In this work, we construct a performance prediction dataset by synthesizing and mapping RTL designs into netlists using Yosys~\cite{yosys} with the SkyWater 130nm technology library~\citep{skywater_pdk}. 
We then utilize open-source ABC~\citep{abc} tool to extract delay and area metrics, where delay metric is reported by the static timing analysis, and area metric reflects the total logic footprint, which directly impacts manufacturing cost.
This process provides a dataset that captures essential performance characteristics of RTL designs, facilitating learning-based performance estimation.
For a comprehensive summary of all dataset statistics, please refer to the Appendix~\ref{appendix:dataset_statistics}.

\section{Methodology}
\label{sec_methodology}

\begin{table*}[htbp]
\centering
\begin{tabular}{l|ccc|ccc}
\hline
\multirow{2}{*}{Model} & \multicolumn{3}{c|}{syntax}                            & \multicolumn{3}{c}{function}                           \\ \cline{2-7} 
                       & pass@1           & pass@5           & pass@10          & pass@1           & pass@5           & pass@10          \\ \hline
GPT-3.5                & 56.50\%          & 69.72\%          & 71.75\%          & 30.10\%          & 39.59\%          & 41.40\%          \\
GPT-4o                 & 72.00\%          & 77.31\%          & 78.53\%          & 49.70\%          & 56.80\%          & 58.84\%          \\
o1-preview             & \textbf{76.20\%} & \textbf{83.71\%} & \textbf{84.00\%} & \textbf{50.00\%} & \textbf{60.86\%} & \textbf{62.52\%} \\ \hline
CodeV-CodeLlama        & 47.70\%          & 74.96\%          & 82.20\%          & 22.00\%          & 39.49\%          & 45.74\%          \\
CodeV-CodeQwen         & 51.50\%          & 77.71\%          & 82.17\%          & 23.10\%          & 44.54\%          & 52.22\%          \\
CodeV-DeepSeek         & 57.60\%          & 80.23\%          & 83.25\%          & 30.00\%          & 49.63\%          & 54.74\%          \\ \hline
DeepRTL-220m           & 60.69\%          & 78.81\%          & 80.88\%          & 28.79\%          & 45.86\%          & 49.66\%          \\
DeepRTL-16b            & 63.79\%          & 74.82\%          & 80.05\%          & 38.91\%          & 47.24\%          & 51.72\%          \\ \hline
Llama-3.1              & 32.40\%          & 57.01\%          & 62.76\%          & 14.60\%          & 26.04\%          & 30.16\%          \\
DeepSeek-Coder         & 59.30\%          & 72.38\%          & 74.67\%          & 31.40\%          & 39.59\%          & 42.57\%          \\ \hline
DeepRTL2$^{1st}$-Direct (Llama)           & 54.48\%          & 63.52\%          & 67.99\%          & 16.28\%          & 28.78\%          & 32.76\%          \\
DeepRTL2$^{1st}$-Direct (DeepSeek)        & 60.60\%          & 73.12\%          & 75.70\%          & 32.50\%          & 44.42\%          & 47.96\%          \\ \hline
DeepRTL2$^{1st}$ (Llama)                  & 67.90\%          & 77.53\%          & 79.52\%          & {\ul 43.70\%}          & 49.98\%          & 50.00\%          \\
DeepRTL2$^{1st}$ (DeepSeek)               & 63.50\%          & 76.74\%          & 80.10\%          & 39.70\%          & 51.96\%          & 54.70\%          \\ \hline
DeepRTL2 (Llama)          & 68.30\%          & {\ul 81.31\%}    & {\ul 83.36\%}    & 33.70\%          & 49.57\%          & 52.90\%          \\
DeepRTL2 (DeepSeek)       & {\ul 71.60\%}    & 80.58\%          & 81.75\%          & 38.50\%    & {\ul 52.62\%}    & {\ul 55.99\%}    \\ \hline
\end{tabular}
\caption{The performance evaluation for RTL code generation using the pass@$k$ metric, with $k$ set to 1, 5, and 10. The best results among all models are bolded, and the best results among open-source models are underlined.}
\vspace{-12pt}
\label{tab:code_generation_result}
\end{table*}
\begin{table}[htbp]
\centering
\begin{tabular}{l|c}
\hline
Model                   & F1
\\ \hline
text-embedding-3-small           & 0.189                         \\
text-embedding-3-large           & 0.290                       \\ \hline
GritLM-7B           & 0.269                         \\ \hline
DeepRTL2$^{no\text{-}hard}$ (Llama)        & \textbf{0.476}          \\ 
DeepRTL2$^{no\text{-}hard}$ (DeepSeek)        & {\ul 0.464}          \\ \hline
DeepRTL2 (Llama)     & 0.463    \\
DeepRTL2 (DeepSeek)     & 0.453     \\ \hline
\end{tabular}
\caption{The performance evaluation for natural language code search using the F1 metric. The best result is bolded, and the second-best result is underscored.}
\vspace{-12pt}
\label{tab:natural_language_code_search_result_main}
\end{table}

\subsection{Model Training}
We choose Llama-3.1~\citep{dubey2024llama} and DeepSeek-Coder~\citep{guo2024deepseek} as the base models for training. Specifically, we fine-tune meta-llama/Llama-3.1-8B-Instruct\footnote{\url{https://huggingface.co/meta-llama/Llama-3.1-8B-Instruct}} and deepseek-ai/deepseek-coder-6.7b-instruct\footnote{\url{https://huggingface.co/deepseek-ai/deepseek-coder-6.7b-instruct}}. 
Our training consists of two stages. In the first stage, we follow the curriculum learning strategy adopted by DeepRTL~\citep{liu2025deeprtl} and train the base model solely on RTL code generation and understanding data. 
In the second stage, we incorporate embedding data into the training set and train the model on both RTL code generation/understanding and embedding tasks, utilizing the training framework of GRIT~\citep{muennighoff2025generative}.
\subsubsection{First-Stage Training}
Following DeepRTL, we apply a curriculum learning strategy in the first stage of our training pipeline, which can be further divided into four sub-stages: training with line-level data, module-level data with specifications, module-level data with high-level descriptions, data with varying prompts. For details on these sub-stages, please refer to Appendix~\ref{appendix:details_of_firststage_training}.
\subsubsection{Second-Stage Training}
Following GRIT, in the second stage of training, we combine the generation/understanding and embedding tasks. 
For the generation/understanding training, we reuse the high-quality data from the fourth sub-stage of the first-stage training. 
For the embedding task, we employ contrastive learning to learn contextualized representations that preserve the semantic information of the original text and code. Details for constructing the contrastive learning training set can be found in Appendix~\ref{appendix:contrastive_learning_training_set_construction}. 
In the embedding part of the second-stage training, we first use data that does not contain hard negatives and then incorporate data with hard negative samples. For more details on the loss functions at different sub-stages, please refer to Appendix~\ref{appendix:training_loss_function}.
For additional details on the hyperparameters and hardware resources used, please refer to Appendix~\ref{appendix:hyperparameters}.

\begin{table*}[htbp]
\centering
\resizebox{\textwidth}{!}{
\begin{tabular}{l|ccccccc}
\hline
Model            & BLEU-4         & ROUGE-1        & ROUGE-2        & ROUGE-L        & METEOR         & Emb. Sim.      & GPT Score      \\ \hline
GPT-3.5          & 3.34           & 28.20          & 10.46          & 25.11          & 20.36          & 0.740          & 0.510          \\
GPT-4o           & 4.59           & 29.26          & 11.48          & 25.74          & 22.78          & 0.761          & 0.549          \\
o1-preview       & 3.73           & 28.00          & 10.39          & 24.98          & 20.48          & 0.748          & 0.535          \\ \hline
CodeV-DeepSeek   & 3.05           & 25.14          & 9.78           & 23.25          & 20.23          & 0.705          & 0.495          \\
CodeV-CodeQwen   & 2.80           & 24.91          & 8.27           & 22.75          & 21.07          & 0.747          & 0.499          \\ \hline
DeepRTL-220m     & 13.06          & 37.56          & 19.85          & {\ul 34.72}          & 34.37          & 0.806          & 0.600          \\
DeepRTL-16b      & 12.85          & 37.43          & 19.34          & 34.63          & 33.09          & 0.802          & 0.597          \\ \hline
Llama-3.1        & 2.68           & 25.37          & 10.39          & 23.75          & 17.16          & 0.730          & 0.430          \\
DeepSeek-Coder   & 2.56           & 24.52          & 7.72           & 22.45          & 22.83          & 0.756          & 0.571          \\ \hline
DeepRTL2$^{1st}$-Direct (Llama)      & 11.28           & 34.29          & 16.35          & 33.63          & 27.73          & 0.754          & 0.580          \\
DeepRTL2$^{1st}$-Direct (DeepSeek)   & 12.07           & 36.37          & 17.78           & 33.78          & 28.56          & 0.767          & 0.602          \\ \hline
DeepRTL2$^{1st}$ (Llama)             & 13.34           & 37.74          & 19.54          & \textbf{34.76}          & 33.46          & 0.798          & 0.594          \\
DeepRTL2$^{1st}$ (DeepSeek)          & 13.53           & 37.52          & 19.68           & 34.68          & 33.28          & {\ul 0.814}          & {\ul 0.612}          \\ \hline
DeepRTL2 (Llama)    & {\ul 13.84}    & \textbf{37.97} & {\ul 20.69}    & 34.42 & \textbf{34.75} & 0.813    & 0.603    \\
DeepRTL2 (DeepSeek) & \textbf{13.96} & {\ul 37.93}    & \textbf{20.73} & 34.34    & {\ul 34.74}    & \textbf{0.820} & \textbf{0.616} \\ \hline
\end{tabular}
}
\caption{The performance evaluation for RTL code understanding. BLEU-4 refers to the smoothed BLEU-4 score, while Emb. Sim. represents the embedding similarity metric. The best results are highlighted in bold, and the second-best results are underscored.}
\label{tab:code_understanding_result}
\vspace{-12pt}
\end{table*}
\subsection{Model Evaluation}

For RTL code generation, we utilize the latest version of the widely adopted RTLLM v2.0 benchmark~\citep{lu2024rtllm}, which contains 50 designs paired with corresponding natural language descriptions and testbenches.
To measure Verilog generation accuracy, we use the pass@$k$ metric, which estimates the proportion of problems that can be solved at least once within $k$ attempts:
\begin{equation}
    \text {pass@}k:=\underset{\text {problems}}{\mathbb{E}}\left[\frac{1-\binom{n-c}{k}}{\binom{n}{k}}\right]
\end{equation}
where $n \geq k$ represents the total number of trials for each problem, and $c$ denotes the number of trails that pass the functional check.
In our experiments, we set $n=20$ to mitigate randomness in results. The pass@$k$ metric is reported for both syntactical and functional accuracy. Following RTLCoder~\citep{liu2024rtlcoder}, we evaluate performance across multiple generation temperatures (0.2, 0.5, and 0.8) and report the best performance across these settings.

For RTL code understanding, we use the benchmark constructed in Section~\ref{subsubsection:rtl_code_understanding}. 
To evaluate the model's performance, we apply both traditional machine translation metrics—BLEU~\citep{papineni2002bleu}, ROUGE~\citep{lin2004rouge}, and METEOR~\citep{banerjee2005meteor}—which primarily assess lexical similarity, as well as the embedding similarity and GPT score metrics introduced in DeepRTL~\cite{liu2025deeprtl}, which focus on semantic similarity.
This combination of evaluation metrics provides a comprehensive assessment of the model’s ability to understand RTL code, capturing both surface-level and deeper, semantic-level understanding.
For further details on how to compute these metrics, please refer to the Appendix~\ref{appendix:understanding_evaluation_metrics}.

For natural language code search, we utilize the benchmark introduced in Section~\ref{data:nlcs}. 
To assess the model's ability to retrieve relevant code from a large codebase based on a user’s query, we follow the bitext mining setting from MTEB~\citep{muennighoff2022mteb}. 
In our evaluation process, the inputs consist of two sets: the first set contains functional descriptions, while the second set consists of Verilog code snippets. 
For each description in the first set, the best matching code snippet in the second set is identified using cosine similarity. 
We report F1 score, precision, and recall for each model, with F1 serving as the primary evaluation metric for natural language code search.

For functionality equivalence checking, we utilize the benchmark introduced in Section~\ref{data:fec}. 
To evaluate the models' ability to check functional equivalence, we follow the pair classification setting from MTEB~\citep{muennighoff2022mteb}. 
In this evaluation, the inputs consist of several pairs of RTL codes. 
For each pair, the model assigns a binary label: 1 for "functionally equivalent" and 0 for "functionally inequivalent".
The binary label is determined by calculating the cosine similarity of their embeddings and comparing the similarity score to a predefined threshold. 
For each model, we first identify the optimal accuracy threshold and compute the accuracy score. 
We then determine the best F1 threshold and report the F1, precision, and recall scores. 
Finally, we calculate the average precision score based on the similarity scores of the code pairs and their corresponding ground-truth labels. Average precision is the primary evaluation metric for RTL code functionality equivalence checking, with other metrics also reported.

For performance prediction, we use the dataset introduced in Section~\ref{data:pp}. 
This task aims to test the expressive power of code embeddings for predicting performance metrics, such as area and delay, at the early stage of RTL design. 
To achieve this, we first encode each code snippet into a fixed-length vector and create a new dataset in the format $\{(\text{code\_embedding}\in\mathbb{R}^p, \text{area}\in\mathbb{R}, \text{delay}\in\mathbb{R})_i\}_{i=1}^n$, where $p$ is  
the embedding dimension and $n$ is the dataset size. 
The dataset is then split into training and test sets at an 80:20 ratio. 
In this paper, we use XGBoost~\citep{chen2016xgboost} as the regression model, training separate models for area and delay prediction.
The trained models are evaluated on the test set using r2\_score, mean absolute percentage error (MAPE) and root relative squared error (RRSE), with their formulas provided below:
\begin{equation}
    \text{r2\_score}(y,\hat{y})=1 - \frac{\sum_{i=1}^n(y_i-\hat{y_i})^2}{\sum_{i=1}^{n}(y_i-\bar{y})^2}
\end{equation}
\begin{equation}
    \text{MAPE}(y,\hat{y})=\frac{1}{n}\sum_{i=1}^{n}|\frac{y_i-\hat{y_i}}{y_i}|\times 100\%
\end{equation}
\begin{equation}
    \text{RRSE}(y,\hat{y})=\sqrt{\frac{\sum_{i=1}^{n}(y_i-\hat{y_i})^2}{\sum_{i=1}^{n}(y_i-\bar{y_i})^2}}
\end{equation}

\section{Experimental Results}
\label{sec_experiment}

\subsection{Generation-Based Tasks}

\begin{table*}[htbp]
\centering
\vspace{-12pt}
\begin{tabular}{l|ccc|ccc}
\hline
\multirow{2}{*}{Model} & \multicolumn{3}{c|}{Area}                            & \multicolumn{3}{c}{Delay}                           \\ \cline{2-7} 
                       & r2\_score           & MAPE           & RRSE          & r2\_score           & MAPE           & RRSE          \\ \hline
text-embedding-3-small     & 0.603         & 5.568          & 0.630          & 0.608          & 0.883          & 0.626         \\
text-embedding-3-large     & 0.699 & 4.446 & 0.548 & 0.699 & 0.705 & 0.548 \\ \hline
GritLM-7B         & 0.651          & 3.878          & 0.591          & 0.651          & 0.726          & 0.591         \\ \hline
DeepRTL2$^{no\text{-}hard}$ (Llama)        & 0.510          & 2.828          & 0.700          & 0.735          & 0.471          & 0.515         \\ 
DeepRTL2$^{no\text{-}hard}$ (DeepSeek)        & \textbf{0.805}          & 2.947          & \textbf{0.445}          & 0.743          & {\ul 0.449}          & 0.507         \\ \hline
DeepRTL2 (Llama) & 0.759 & {\ul 1.966} & 0.490 & \textbf{0.773} & 0.469 & \textbf{0.476} \\
DeepRTL2 (DeepSeek) & {\ul 0.773} & \textbf{1.598} & {\ul 0.476} & {\ul 0.772} & \textbf{0.448} & {\ul 0.478} \\
\hline
\end{tabular}
\caption{The performance evaluation for performance prediction on area and delay using r2\_score, MAPE and RRSE metrics. The best results among all models are bolded, and the second-best results are underscored.}
\vspace{-12pt}
\label{tab:performance_prediction_result}
\end{table*}

\begin{table}[ht]
\centering
\resizebox{0.48\textwidth}{!}{
\begin{tabular}{l|c}
\hline
Model                   & Average Precision
\\ \hline
text-embedding-3-small           &   0.565                       \\
text-embedding-3-large           &   0.498                     \\ \hline
GritLM-7B           &    0.541                      \\ \hline
DeepRTL2$^{no\text{-}hard}$ (Llama)        & 0.518          \\ 
DeepRTL2$^{no\text{-}hard}$ (DeepSeek)        & 0.481          \\ \hline
DeepRTL2 (Llama)     &  \textbf{0.667}   \\
DeepRTL2 (DeepSeek)     &   {\ul 0.591}    \\ \hline
\end{tabular}
}
\caption{The performance evaluation for RTL code functionality equivalence checking using the average precision metric. The best result among all models is bolded, and the second-best result is underscored.}
\label{tab:functional_equivalence_checking_result_main}
\vspace{-12pt}
\end{table}

For comparison, we select several baseline models: the state-of-the-art commercial models, OpenAI's GPT-3.5, GPT-4o, and o1-preview, which represent the most advanced general-purpose LLMs currently available.
We also include the CodeV series~\citep{zhao2024codev}, a collection of leading open-source models specifically designed for RTL code generation, as well as the original DeepRTL models~\cite{liu2025deeprtl}, which have shown strong performance in both RTL code generation and understanding.
All these models have demonstrated excellent capabilities in Verilog generation-based tasks~\citep{liu2025deeprtl}, making them strong baselines for evaluating the performance of DeepRTL2. 
Additionally, we report the performance of base models, Llama-3.1 and DeepSeek-Coder, to show the effectiveness of our dataset construction and training strategy.

Table~\ref{tab:code_generation_result} reports the pass@$k$ results for RTL code generation across different models, with $k$ set to 1, 5, and 10.
The results show that o1-preview outperforms all other models, likely due to its design for addressing complex tasks, including programming.
The DeepRTL2 models, however, achieve the best performance among all open-source models, with results comparable to GPT-4o. The performance improvement from base models to DeepRTL2 highlights the effectiveness of our dataset construction process and training strategy.
Furthermore, DeepRTL2 outperforms the original DeepRTL models, likely due to the incorporation of additional open-source datasets, aside from data sourced from GitHub, and the inclusion of more diverse problem formulations that enhance DeepRTL2's generalization ability.
Given that DeepRTL2 is a multi-task model and the generation benchmark may overlap with the training data used by OpenAI's models, these results highlight DeepRTL2’s impressive performance for this task.

Table~\ref{tab:code_understanding_result} presents the results for RTL code understanding. Since the CodeV-CodeLlama model outputs random messages for this task, we exclude it from the comparison. The results show that DeepRTL2 models significantly outperform all other models, including the previous state-of-the-art DeepRTL models, underscoring its strong capabilities in RTL code understanding.
Notably, DeepRTL2 surpasses GPT-4o by a substantial margin, despite the fact that its training data is annotated using GPT-4o. The main reason is that during benchmark testing, all models, including GPT-4o, are required to generate high-level functional descriptions directly from RTL code. As shown in Appendix~\ref{appendix:human_evaluation_for_generated_annotations}, CoT-based annotations are more accurate than direct annotations. This enhanced annotation quality contributes to DeepRTL2's superior performance in RTL code understanding.

\subsection{Embedding-Based Tasks}
Since none of the existing models are specifically designed for RTL embedding-based tasks, the baselines used for the generation-based tasks, \emph{e.g.}, CodeV series and DeepRTL models, perform poorly in this setting.
These models show near-zero performance, with an F1 score close to 0 on the natural language code search task and an average precision of approximately 0.5 on the functionality equivalence checking task.
Therefore, we select state-of-the-art general-purpose embedding models as baselines for comparison.
These include OpenAI's text embedding models (text-embedding-3-small, text-embedding-3-large)~\citep{neelakantan2022text} and open-source models like GritLM-7B~\citep{muennighoff2025generative}.

Table~\ref{tab:natural_language_code_search_result_main} presents the F1 scores for the natural language code search task. The results show that our DeepRTL2 models outperform all baseline models by a significant margin, demonstrating the effectiveness of our dataset and training strategy for this task. For the full evaluation results on natural language code search, please refer to Appendix~\ref{appendix:full_evaluation_results}.

Table~\ref{tab:functional_equivalence_checking_result_main} presents the average precision scores for the functionality equivalence checking task. The results show that DeepRTL2 models outperform all other baselines, demonstrating their effectiveness in capturing functional relationships between RTL modules. The full evaluation results are in Appendix~\ref{appendix:full_evaluation_results}.
It is important to emphasize that our embedding-based verification is not intended to replace the traditional verification process, but rather to serve as an efficient preliminary step that can significantly streamline the verification flow.

Table~\ref{tab:performance_prediction_result} presents the results for performance prediction on area and delay. Our DeepRTL2 series models outperform the baseline models across all metrics. These results highlight that the code embeddings generated by the DeepRTL2 models are more expressive for predicting performance-related metrics such as area and delay.

\subsection{Ablation Studies}

In this section, we conduct ablation studies to demonstrate the effectiveness of different dataset components and training strategies. 
In the first training stage, we adopt a curriculum learning strategy, where the model is progressively trained on line-level data, module-level data with specifications, module-level data with high-level descriptions, and data with varying prompts. While the benefits of curriculum learning have been shown in DeepRTL~\citep{liu2025deeprtl}, we extend this analysis with additional comparisons. Specifically, we compare our first-stage model (DeepRTL2$^{1st}$) with a variant trained without curriculum learning (DeepRTL2$^{1st}$-Direct), both focused on generation-based tasks. As shown in Table~\ref{tab:code_generation_result} and Table~\ref{tab:code_understanding_result}, the incorporation of curriculum learning significantly improves performance for both code generation and understanding tasks. When we further introduce the second-stage training, \textit{i.e.}, GRIT-based fine-tuning, the performance improves even more, demonstrating the effectiveness of both curriculum learning and GRIT-based fine-tuning strategies.

In the second training stage, we combine contrastive learning and curriculum learning to ensure that our model performs effectively on embedding-based tasks. Specifically, we start with data that excludes hard negatives and gradually introduce hard negative samples, which improves overall performance. To evaluate this strategy, we compare DeepRTL2 with and without hard negatives (DeepRTL2$^{no\text{-}hard}$) in Tables~\ref{tab:natural_language_code_search_result_main}, \ref{tab:performance_prediction_result}, and \ref{tab:functional_equivalence_checking_result_main}. Since hard negatives primarily influence contrastive learning, these comparisons focus on embedding-based tasks, with negligible impact on generation-based performance. 
The results show a minor drop in natural language code search accuracy but substantial gains in functionality equivalence checking and performance prediction. Despite the small accuracy decrease in the natural language code search task, DeepRTL2 still outperforms powerful baseline embedding models. This improvement in functionality equivalence checking and performance prediction justifies our decision to integrate hard negatives into the training process.

\section{Conclusion}
\label{sec_conclusion}
In this work, we present DeepRTL2, a novel family of LLMs that unifies both generation- and embedding-based tasks at the RTL stage, offering a comprehensive solution to the diverse challenges in EDA. 
By addressing critical tasks including RTL code generation, understanding, natural language code search, functionality equivalence checking, and performance prediction, DeepRTL2 significantly improves the efficiency of hardware design workflows.
To develop DeepRTL2, we have curated a comprehensive dataset and established new benchmarks specifically designed for these tasks, particularly the embedding-based ones, for which no suitable resources previously existed.
Furthermore, we adapt the GRIT approach to fine-tune the model, enabling it to manage both generation- and embedding-based tasks effectively. 
Extensive experimentation demonstrates that DeepRTL2 achieves state-of-the-art performance across all evaluated tasks, advancing the application of LLMs in hardware design.
\section{Limitations}
\label{sec_limitation}

There are two main limitations in our work. First, due to the multi-task nature of our model and constraints in time and computing resources, we may not have employed the most optimal training strategy and hyperparameter settings to maximize performance across all tasks.
Second, performance prediction directly at the RTL stage is challenging, as RTL designs typically lack detailed information about delay and area metrics. Although our model outperforms others in the evaluation, a significant gap remains in achieving accurate predictions.
We hypothesize that incorporating the control data flow graph (CDFG) of RTL designs, which offers a more structured representation of the design's behavior, may facilitate better learning of performance characteristics.
In future work, we plan to explore how incorporating CDFG into the DeepRTL2 model series could improve the model’s ability to predict performance metrics more accurately.

\bibliography{custom}

\appendix

\section{Human Evaluation of Generated Annotations}
\label{appendix:human_evaluation_for_generated_annotations}
To evaluate the reliability and accuracy of GPT-4o-generated annotations, we conduct a human evaluation focusing primarily on the accuracy of high-level functional descriptions, as this is the most challenging and critical aspect of the generation-based tasks. We randomly sample 200 annotated RTL modules and ask professional hardware designers to verify the correctness of the generated descriptions. The human evaluation results show that approximately 90\% of these annotations are accurate. In comparison, when we test direct annotations, \emph{i.e.}, generating high-level functional descriptions directly from the original code, the accuracy drops significantly to 70\%. This significant difference further demonstrates the effectiveness of the CoT-based annotation strategy.

Additionally, GPT-4o is employed for rewriting RTL code in the functionality equivalence checking task. For this task, we address concerns about accuracy by using EDA tools to verify the functionality equivalence of the rewritten code against the original code. 
\begin{figure}[ht]
    \centering
    \includegraphics[width=\linewidth]{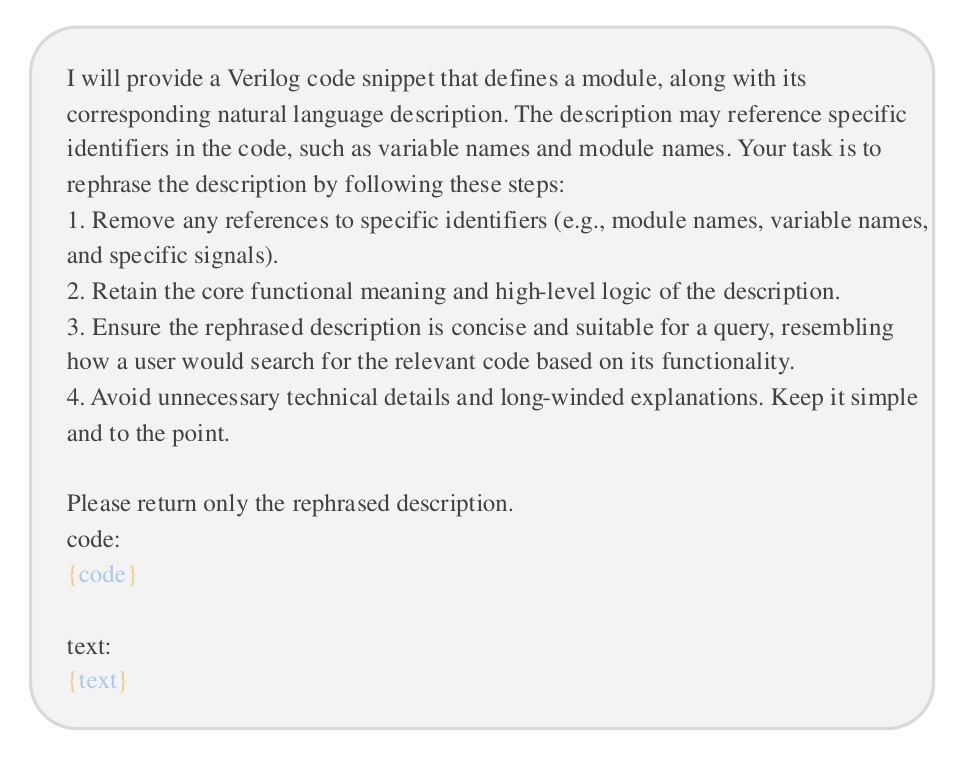}
    \vspace{-12pt}
    \caption{The instruction for rephrasing the code description into the user query format.}
    \vspace{-12pt}
    \label{fig:rephrase_description}
\end{figure}
Therefore, all the data collected for this task is validated as ground truth, ensuring the quality and correctness of the rewritten RTL code.

\section{Prompt For Rephrasing Descriptions}
\label{appendix:prompt_for_rephrasing_descriptions}

Figure~\ref{fig:rephrase_description} shows the instruction given to GPT-4o to rephrase the code descriptions into their corresponding user query formats.

\section{Code Rewrite Instructions}
\label{appendix:code_rewrite_instructions}

\begin{figure*}[ht]
    \centering
    \includegraphics[width=0.9\linewidth]{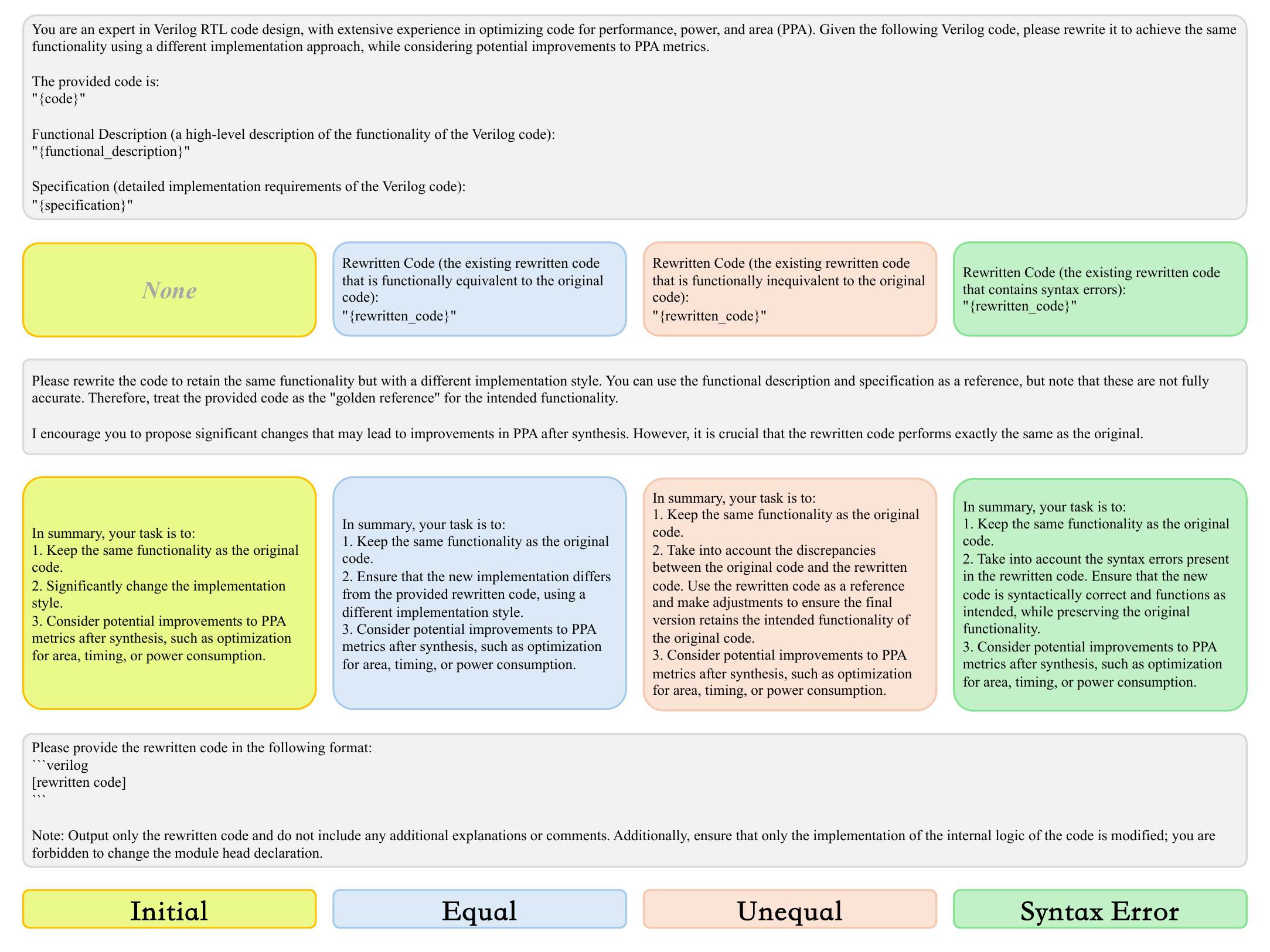}
    \vspace{-5pt}
    \caption{The code rewrite instructions used to construct the functionality equivalence checking dataset.}
    \vspace{-12pt}
    \label{fig:code_rewrite_instructions}
\end{figure*}

Figure~\ref{fig:code_rewrite_instructions} illustrates the code rewrite instructions provided to GPT-4o for constructing the functionality equivalence checking dataset.
The leftmost column presents the instruction used during the initial rewrite process, where only the original RTL code is available.
The subsequent three columns represent instructions based on previously rewritten code, corresponding to the following cases: (1) equivalent rewritten code, (2) inequivalent rewritten code, and (3) rewritten code with syntax errors.
Notably, in addition to the code itself, we also include the functional description and specification from Section~\ref{subsubsection:rtl_code_generation}. This additional context helps the model better understand the intended functionality, leading to improved accuracy in rewriting the code while preserving its functionality.

\section{Dataset Statistics}
\label{appendix:dataset_statistics}
\begin{table*}[htbp]
\centering
\resizebox{\textwidth}{!}{
\begin{tabular}{l|l|l|c}
\hline
Task                                                & Description                                            & Source       & Count  \\ \hline
\multirow{8}{*}{RTL Code Generation/Understanding}  & Line Level                                             & DeepRTL2        & 341310 \\ \cline{2-4} 
                                                    & \multirow{3}{*}{Module Level (Detailed Specification)} & DeepRTL2        & 45519  \\ \cline{3-4} 
                                                    &                                                        & MG-Verilog   & 10035  \\ \cline{3-4} 
                                                    &                                                        & DeepCircuitX & 32809  \\ \cline{2-4} 
                                                    & \multirow{4}{*}{Module Level (High-Level Description)} & DeepRTL2        & 46876  \\ \cline{3-4} 
                                                    &                                                        & RTLCoder     & 25001  \\ \cline{3-4} 
                                                    &                                                        & MG-Verilog   & 10037  \\ \cline{3-4} 
                                                    &                                                        & DeepCircuitX & 38179  \\ \hline
Natural Language Code Search                        & N/A                                                    & DeepRTL2        & 59700       \\ \hline
\multirow{2}{*}{Functionality Equivalence Checking} & Equal Pairs                                            & DeepRTL2        & 9532   \\ \cline{2-4} 
                                                    & Unequal Pairs                                          & DeepRTL2        & 23330  \\ \hline
\multirow{2}{*}{Performance Prediction}             & Area                                                   & DeepRTL2        & 18766       \\ \cline{2-4} 
                                                    & Delay                                                  & DeepRTL2        & 18766       \\ \hline
\end{tabular}
}
\caption{The overall dataset statistics for all evaluated tasks.}
\label{tab:dataset_statistics}
\end{table*}
Table~\ref{tab:dataset_statistics} presents the overall statistics for all datasets used across the evaluated tasks. 
Except the performance prediction datasets, all datasets listed in this table are utilized for model training. 
For the performance prediction datasets, we split them in an 80:20 ratio, creating a training set with 15,000 samples and a test set with 3,766 samples.
For performance prediction, we regress area and delay based on code embeddings, without tuning the model.

\section{Details of First-Stage Training}
\label{appendix:details_of_firststage_training}
In Section~\ref{subsubsection:rtl_code_generation}, we construct a dataset consisting of Verilog modules enriched with line-level comments, detailed specifications, and succinct high-level functional descriptions.
These three levels of annotations correspond to the first three sub-stages of our first-stage training pipeline.
In the first sub-stage, we train the model using line-level data, where each line of Verilog code is paired with a corresponding natural language comment.
The second sub-stage utilizes module-level data with specifications, providing more detailed descriptions of the Verilog modules. The third sub-stage focuses on module-level data with high-level functional descriptions, offering a broader functional overview of the code.
To further refine the dataset and adapt it to a wider range of scenarios, we introduce a fourth sub-stage, where GPT-4o generates varying prompts based on the high-quality data from the third sub-stage.
These varying prompts represent different problem descriptions used to generate Verilog code. We find that incorporating this sub-stage improves the model's performance and robustness, as it allows the model to better generalize across a wide range of code generation tasks.

\section{Contrastive Learning Training Set Construction}
\label{appendix:contrastive_learning_training_set_construction}
In the second-stage training, we apply contrastive learning to enable the model to (1) determine whether a Verilog module matches a given functional description; and (2) assess whether two Verilog code snippets are functionally equivalent.

To construct a dataset for contrastive learning, we first prompt GPT-4o to rewrite Verilog code snippets from the natural language code search training set. The rewrite process is illustrated in Figure~\ref{fig:code_rewrite_process}. After several iterations, we combine the original natural language code search training set with their rewritten code snippets, resulting in four types of new data samples:
\begin{itemize}
    \item type a: \{original\_text, original\_code\}
    \item type b: \{original\_text, original\_code, equivalent\_code\}
    \item type c: \{original\_text, original\_code, inequivalent\_code\}
    \item type d: \{original\_text, original\_code, equivalent\_code, inequivalent\_code\}
\end{itemize}
Since the format of an original data sample in the natural language code search training set is \{original\_text, original\_code\}, the four types of data samples correspond to the following scenarios:
\begin{itemize}
    \item type a corresponds to the case where all rewritten code snippets contain syntax errors.
    \item type b corresponds to the case where all rewritten code snippets, free of syntax errors, are functionally equivalent to the original code.
    \item type c corresponds to the case where all rewritten code snippets, free of syntax errors, are not functionally equivalent to the original code.
    \item type d corresponds to the case where some rewritten code snippets, free of syntax errors, are functionally equivalent to the original code, while others are not.
\end{itemize}
For all four types of data samples, we convert them into contrastive learning samples as follows:
\begin{itemize}
    \item type a:
    \begin{itemize}
        \item \textcolor{blue}{\{"query": original\_code, "pos": original\_text, "neg": None\}}
        \item \textcolor{blue}{\{"query": original\_text, "pos": original\_code, "neg": None\}}
    \end{itemize}
    \item type b:
    \begin{itemize}
        \item \textcolor{blue}{\{"query": original\_code, "pos": original\_text, "neg": None\}}
        \item \textcolor{blue}{\{"query": original\_text, "pos": original\_code, "neg": None\}}
        \item \textcolor{blue}{\{"query": original\_code, "pos": equivalent\_code, "neg": None\}}
        \item \textcolor{blue}{\{"query": equivalent\_code, "pos": original\_code, "neg": None\}}
    \end{itemize}
    \item type c:
    \begin{itemize}
        \item \textcolor{purple}{\{"query": original\_code, "pos": original\_text, "neg": inequivalent\_code\}}
        \item \textcolor{blue}{\{"query": original\_text, "pos": original\_code, "neg": None\}}
    \end{itemize}
    \item type d:
    \begin{itemize}
        \item \textcolor{purple}{\{"query": original\_code, "pos": original\_text, "neg": inequivalent\_code\}}
        \item \textcolor{purple}{\{"query": original\_code, "pos": equivalent\_code, "neg": inequivalent\_code\}}
        \item \textcolor{blue}{\{"query": original\_text, "pos": original\_code, "neg": None\}}
        \item \textcolor{purple}{\{"query": equivalent\_code, "pos": original\_code, "neg": inequivalent\_code\}}
    \end{itemize}
\end{itemize}
In each of the contrastive learning samples above, the key ``pos'' refers to the positive instance of the query code/text, while the key ``neg'' refers to the hard negative instance. In the embedding part of the second-stage training, we first use samples colored \textcolor{blue}{blue} that do not contain hard negatives and then incorporate samples colored \textcolor{purple}{purple} with hard negative instances.

\section{Training Loss Function}
\label{appendix:training_loss_function}
\begin{table*}[htbp]
\centering
\begin{tabular}{l|ccc}
\hline
Model            & Precision         & Recall        & F1 (Main Metric)                    \\ \hline
text-embedding-3-small          &  0.173          & 0.241          & 0.189                              \\
text-embedding-3-large           & 0.273           & 0.340          & 0.290                              \\ \hline
GritLM-7B       & 0.255           & 0.320          & 0.269                              \\ \hline
DeepRTL2$^{no\text{-}hard}$ (Llama)        & \textbf{0.469}          & \textbf{0.497}          & \textbf{0.476}          \\ 
DeepRTL2$^{no\text{-}hard}$ (DeepSeek)        & {\ul 0.456}          & 0.489          & {\ul 0.464}          \\ \hline
DeepRTL2 (Llama)       & 0.450 & {\ul 0.493} & 0.463  \\
DeepRTL2 (DeepSeek) & 0.443 & 0.481    & 0.453      \\ \hline
\end{tabular}
\caption{The full performance evaluation results for natural language code search. The best results among all models are bolded, and the second-best results are underscored.}
\label{tab:natural_language_code_search_result_full}
\vspace{-8pt}
\end{table*}
In the second stage of training, we combine generation/understanding and embedding tasks. For generation/understanding, we reuse high-quality data from the fourth sub-stage of the first training stage. For the embedding tasks, we apply contrastive learning to learn contextualized representations that preserve the semantic information of text and code. In the embedding part of the second-stage training, we first use data without hard negatives and later incorporate data with hard negatives. The embedding loss function is defined as follows:

\begin{equation}
    E_i^+ = \exp \left( \frac{\sigma(f_{\theta}(x_i), f_{\theta}(x_i^+))}{\tau} \right)
\end{equation}

\begin{equation}
    S_i^+ = \sum_{j=1}^{M} \exp \left( \frac{\sigma(f_{\theta}(x_i), f_{\theta}(x_j^+))}{\tau} \right)
\end{equation}

\begin{equation}
    S_i^- = \sum_{j=1}^{M} \exp \left( \frac{\sigma(f_{\theta}(x_i), f_{\theta}(x_j^-))}{\tau} \right)
\end{equation}

\begin{equation}
    \mathcal{L}_{emb1} = -\frac{1}{M} \sum_{i=1}^{M} \log \left( \frac{E_i^+}{S_i^+} \right)
\end{equation}

\begin{equation}
    \mathcal{L}_{emb2} = -\frac{1}{M} \sum_{i=1}^{M} \log \left( \frac{E_i^+}{S_i^+ + S_i^-} \right)
\end{equation}

\begin{table*}[htbp]
\centering
\vspace{-12pt}
\resizebox{\textwidth}{!}{
\begin{tabular}{l|ccccc}
\hline
Model            & Average Precision (Main Metric)         & Accuracy        & F1        & Precision   & Recall           \\ \hline
text-embedding-3-small          &  0.565  &  0.581  &  0.646     & 0.525 & 0.840                 \\
text-embedding-3-large           &  0.498   &  0.544   &  0.647     &  0.478 & \textbf{1.000}   \\ \hline
GritLM-7B       &  0.541   &  0.613     & 0.661      &  0.503  & 0.960            \\ \hline
DeepRTL2$^{no\text{-}hard}$ (Llama)        & 0.518          & 0.594          & 0.661    & 0.497 & {\ul 0.987}         \\ 
DeepRTL2$^{no\text{-}hard}$ (DeepSeek)        & 0.481          & 0.581          & 0.658    & 0.497 & 0.973         \\ \hline
DeepRTL2 (Llama)       & \textbf{0.667} & \textbf{0.681} & \textbf{0.723} & \textbf{0.575} & 0.973 \\
DeepRTL2 (DeepSeek) & {\ul 0.591} & {\ul 0.619}  & {\ul 0.708} & {\ul 0.552} & {\ul 0.987}   \\ \hline
\end{tabular}
}
\vspace{-6pt}
\caption{The full performance evaluation results for RTL code functionality equivalence checking. The best results among all models are bolded, and the second-best results are underscored.}
\vspace{-12pt}
\label{tab:functional_equivalence_checking_result_full}
\end{table*}

where $M$ is the batch size, $x_i$ is the $i$-th training sample, $f_\theta$ is the embedding function (in this paper, we use position-weighted mean pooling method introduced in SGPT~\citep{muennighoff2022sgpt} to obtain sentence embeddings), $\tau$ is the temperature hyperparameter, and $\sigma$ is the similarity function (typically cosine similarity).
$x_i^+$ is the positive instance of the $i$-th training sample, while $x_i^-$ is the hard negative of the $i$-th training sample.
$\mathcal{L}_{emb1}$ represents the embedding loss when no hard negative is available for each training sample, while $\mathcal{L}_{emb2}$ corresponds to the embedding loss when a hard negative instance is present for each sample.

For generation/understanding, we adopt the traditional next-token cross-entropy loss:
\begin{equation}
   \mathcal{L}_{gen}=-\frac{1}{N}\sum_{i=1}^{N}\log P(f_{\theta,\eta}(x^{(i)})|f_{\theta,\eta}(x^{(<i)})) 
\end{equation}
where $\eta$ is the language modeling head used for generation-based tasks. In the second-stage training, we first use $\mathcal{L}_1=\mathcal{L}_{emb1}+\mathcal{L}_{gen}$ as the loss function, then switch to $\mathcal{L}_2=\mathcal{L}_{emb2}+\mathcal{L}_{gen}$.

\section{Hyperparameters}
\label{appendix:hyperparameters}
All experiments are conducted on a cluster equipped with eight NVIDIA A800 GPUs, each with 80GB of memory.
Tables~\ref{tab:hyperparameters_first_stage} and \ref{tab:hyperparameters_second_stage} present the hyperparameter settings used in the first-stage and second-stage training, respectively.

\begin{table}[ht]
\centering
\resizebox{0.4\textwidth}{!}{
\begin{tabular}{l|c}
\hline
Hyperparameter Name                   & Value
\\ \hline
finetuning\_type & lora \\
per\_device\_train\_batch\_size & 4 \\
gradient\_accumulation\_steps & 4 \\
lr\_scheduler\_type & cosine \\
warm\_up\_ratio & 0.1 \\
learning\_rate & 5e-5 \\
epochs & 3 \\ \hline

\end{tabular}
}
\vspace{-6pt}
\caption{Hyperparameters selected for the first training stage of DeepRTL2.}
\label{tab:hyperparameters_first_stage}
\vspace{-6pt}
\end{table}

\begin{table}[ht]
\centering
\resizebox{0.4\textwidth}{!}{
\begin{tabular}{l|c}
\hline
Hyperparameter Name                   & Value
\\ \hline
finetuning\_type & full \\
per\_device\_embedding\_batch\_size & 4 \\
per\_device\_generative\_batch\_size & 4 \\
gradient\_accumulation\_steps & 8 \\
lr\_scheduler\_type & linear \\
warmup\_ratio & 0.03 \\
learning\_rate & 2e-5 \\
epochs & 1 \\
temperature ($\tau$) & 0.02 \\ \hline

\end{tabular}
}
\caption{Hyperparameters selected for the second training stage of DeepRTL2.}
\vspace{-6pt}
\label{tab:hyperparameters_second_stage}
\vspace{-12pt}
\end{table}

\section{Understanding Evaluation Metrics}
\label{appendix:understanding_evaluation_metrics}

To evaluate the models' understanding capabilities of RTL code, we apply both traditional machine translation metrics—BLEU~\citep{papineni2002bleu}, ROUGE~\citep{lin2004rouge}, and METEOR~\citep{banerjee2005meteor}—which primarily assess lexical similarity, as well as the embedding similarity and GPT score introduced in DeepRTL~\citep{liu2025deeprtl}, which focus on semantic similarity.
These metrics measure the similarity between the generated descriptions and the ground truth summaries.

Specifically, BLEU measures the proportion of n-grams (sequences of n words) in the generated text that also appear in the reference text.
It calculates the overlap of n-grams (typically up to a length of 4), with higher scores assigned to more matches.
BLEU is precision-focused and rewards the accurate use of words or phrases in the generated descriptions.
In our evaluation, we report the smoothed BLEU-4 score to address zero counts in higher-order n-grams, which helps to avoid penalizing models for small discrepancies.

ROUGE is a recall-based metric that evaluates the proportion of n-grams in the reference summary that are present in the generated summary.
For our evaluation, we report ROUGE-1 (unigram overlap), ROUGE-2 (bigram overlap), and ROUGE-L (longest common subsequence).

METEOR combines both precision and recall while also accounting for synonymy, stemming, and word order. It computes unigram precision and recall and applies a penalty for word order mismatches.
For calculating these traditional machine translation metrics, we directly use the corresponding functions from Python libraries nltk (for BLEU and METEOR) and rouge (for ROUGE).

In contrast to the lexical metrics, embedding similarity and GPT score evaluate semantic similarity by assessing how well the generated description captures the underlying meaning of the RTL code, rather than focusing solely on surface-level word matches.
Embedding similarity computes the cosine similarity between the embeddings of the generated description and the ground truth summary, derived from OpenAI's text-embedding-3-large model. This metric rewards models for producing descriptions that are semantically closer to the reference, even if the wording differs.
The GPT score, based on GPT-4o, quantifies the semantic coherence between descriptions by assigning a similarity score between 0 and 1, where 1 indicates perfect alignment.
Unlike lexical metrics, the GPT score focuses on semantic accuracy rather than exact word matching.
For the prompt used in calculating the GPT score, please refer to DeepRTL.

Together, these metrics offer a comprehensive evaluation of both lexical precision and semantic accuracy, providing a holistic view of the model’s understanding of RTL code.

\section{Full Evaluation Results}
\label{appendix:full_evaluation_results}
\subsection{Natural Language Code Search}
The full evaluation results for natural language code search are presented in Table~\ref{tab:natural_language_code_search_result_full}. Results show that the DeepRTL2 models significantly outperform all baseline models across all metrics. Specifically, DeepRTL2 (Llama) and DeepRTL2 (DeepSeek) achieve F1 scores of 0.463 and 0.453, respectively, surpassing the best baseline model, GritLM-7B, which scores 0.269. 
The higher precision and recall scores for the DeepRTL2 models indicate that they are more effective at retrieving relevant code snippets based on user queries, highlighting the strength of our dataset and training framework. 
These results confirm that DeepRTL2 excels in natural language code search, demonstrating its superior ability to handle hardware-specific queries compared to the baseline models.
\subsection{Functionality Equivalence Checking}
The full evaluation results for RTL code functionality equivalence checking are presented in Table~\ref{tab:functional_equivalence_checking_result_full}.
Results show that the DeepRTL2 models outperform all baseline models across all metrics.
Specifically, DeepRTL2 (Llama) achieves the highest performance with an average precision score of 0.667, F1 score of 0.723, and accuracy of 0.681. In comparison, the best-performing baseline model, GritLM-7B, achieves an average precision of 0.541, an F1 score of 0.661, and accuracy of 0.613. Moreover, DeepRTL2 (DeepSeek) also shows strong performance, with an average precision of 0.591 and an F1 score of 0.708. The significantly higher precision and recall scores for DeepRTL2 models indicate their superior capability in identifying functionally equivalent RTL code compared to the baseline models. These results confirm that DeepRTL2 excels in functionality equivalence checking, demonstrating its effectiveness in hardware-specific tasks over general-purpose models.

\end{document}